# EdgeLLM: A Highly Efficient CPU-FPGA Heterogeneous Edge Accelerator for Large Language Models

Mingqiang Huang, Ao Shen, Kai Li, Haoxiang Peng, Boyu Li, Yupeng Su, Hao Yu*

*Abstract-* The rapid advancements in artificial intelligence (AI), particularly the Large Language Models (LLMs), have profoundly affected our daily work and communication forms. However, it is still a challenge to deploy LLMs on resource-constrained edge devices (such as robots), due to the intensive computation requirements, heavy memory access, diverse operator types and difficulties in compilation. In this work, we proposed EdgeLLM to address the above issues. Firstly, focusing on the computation, we designed mix-precision processing element array together with group systolic architecture, that can efficiently support both FP16*FP16 for the MHA block (Multi-Head Attention) and FP16*INT4 for the FFN layer (Feed-Forward Network). Meanwhile specific optimization on log-scale structured weight sparsity, has been used to further increase the efficiency. Secondly, to address the compilation and deployment issue, we analyzed the whole operators within LLM models and developed a universal data parallelism scheme, by which all of the input and output features maintain the same data shape, enabling to process different operators without any data rearrangement. Then we proposed an end-to-end compiler to map the whole LLM model on CPU-FPGA heterogeneous system (AMD Xilinx VCU128 FPGA). The accelerator achieves 1.91× higher throughput and 7.55× higher energy efficiency than the commercial GPU (NVIDIA A100-SXM4-80G). When compared with state-of-the-art FPGA accelerator of FlightLLM, it shows 10-24% better performance in terms of HBM bandwidth utilization, energy efficiency and LLM throughput.

*Index Terms*—Large Language Model, AI Accelerator

## I. Introduction

Artificial Intelligence has captured keen interest and worldwide attention in the past ten years [1-5]. This is mainly associated with three pivotal advancements: First is the exponential escalation in computational prowess, which mainly attributable to GPU; Second is the evolution of AI algorithm [4-6], especially the convolutional neural networks for computer vision (CV), and the recurrent neural networks for natural language processing (NLP); Third is the AI big data [7,8], which serves as the foundational support for the training of AI models, fueling their precision and efficacy. These factors have coalesced to drive a transformative era in AI.

This work was supported by STI 2030-Major Projects (2022ZD0210600), National Natural Science Foundation of China (NSFC) (Grant No. 92464102, 62034007), Natural Science Foundation of Guangdong Province (Grant 2023B1515020051), Shenzhen Science and Technology Program (Grant No. JCYJ20200109115210307). (Corresponding author: Hao Yu)

Mingqiang Huang and Ao Shen contribute equally in this work.
Mingqiang Huang is with Shenzhen Institute of Advanced Technology, Chinese Academy of Sciences, Shenzhen 518055, China.
Ao Shen, Kai Li, Haoxiang Peng, Boyu Li, Yupeng Su and Hao Yu are with School of Microelectronics, Southern University of Science and Technology, Shenzhen 518055, China.(yuh3@sustech.edu.cn)

Since 2017, Transformer algorithm has rapidly become a dominant force in the field of artificial intelligence in terms of both NLP and CV research field [4-6]. By introducing the self-attention mechanism, Transformer altered traditional sequence modeling methods, no longer relying on convolutional neural networks (CNNs) or recurrent neural networks (RNNs) for feature extraction and sequence handling. This innovation not only enhanced the parallelization capabilities of models, making the training of large-scale models possible, but also significantly improved model performance.

Following the success of Transformer, researchers began to explore even larger models, which typically have billions or more parameters, known as 'large language models' [9-11]. For example, the GPT (Generative Pre-trained Transformer) series of models from OpenAI, has showcased ability to understand and generate high-quality human language [11]. Nowadays, the large language models have begun integrating with other modalities such as vision and audio, showing capable of outstanding performance in cross-domain task. This interactive capability makes AI assistants, chatbots, and virtual agents more human-like and efficient [12-15]. In summary, large language models not only enhance AI's language processing capabilities but also advance broader technological progress in AI, bringing revolutionary changes to industries such as education, entertainment, healthcare, and business.

Nonetheless, the prevailing trend in contemporary AI accelerator architectures, especially those geared towards the demands of large language models, predominantly relies on the GPU paradigm. This conventional design, while powerful, exhibits inherent limitations that render it less than ideal for deployment on edge devices. The constraints imposed by the GPU architecture, such as high power consumption and extensive computational overhead, pose significant challenges to the practicality and efficiency of implementing AI capabilities at the network's periphery. Despite its attributes of high flexibility and high-performance capabilities in specific tasks, FPGA, a crucial component of heterogeneous computing, receives scant mention. The FPGA based implementation of LLM holds considerable promise for substantially enhancing computational efficiency and performance [16-20]. However, several critical challenges remain to be effectively addressed.

The first challenge within LLM is the heavy computation and memory access [21, 22]. Large language models often contain billions of parameters in matrix-matrix multiplication (MatMUL) or vector-matrix multiplication (VMM). We often refer to a model as LLM-6B or LLM-7B, where 6B or 7B denotes the number of weight parameters in the matrix multiplications. The sheer volume of parameters translates into extensive memory requirements, making it difficult to fit these models into the memory of standard hardware. To address such issue, several strategies have been developed, such as model pruning and quantization. Existing solutions typically quantize these weight parameters to INT4 format [21]. However, to ensure computational accuracy, the activation functions in the model remain in FP16 format. Therefore, the system requires matrix multiplications using FP16*INT4 format in the FFN(Feed-Forward Neural Network) layers. Besides, despite its theoretical advantages, sparse representation techniques have yet to gain widespread traction within the industry [23,24]. As such, there exists a clear opportunity of specialized hardware components tailored to the unique requirements of both sparse and INT4 quantized models. On the other hand, in the MHA (Multi-Head Attention) block, KVcache is dynamically generated as activation data, thus the matrix multiplication involving KVcache requires FP16*FP16 format. Therefore, it is highly desired to design mix-precision computation elements for the acceleration of LLMs.

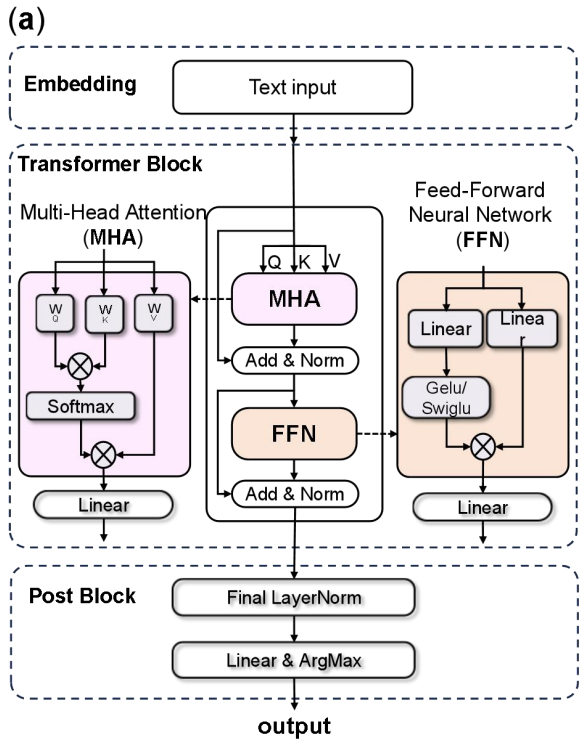

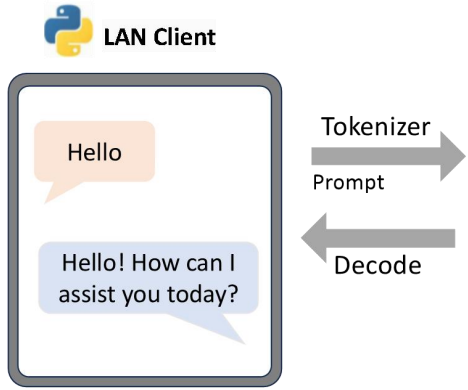
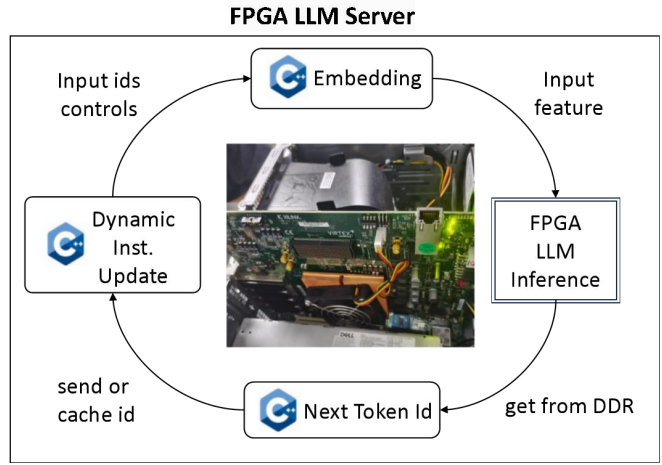

**Fig. 1**. (a) Architecture of the Transformer network and challenges in hardware design for Transformer-like accelerator. (b) Proposed CPU-FPGA heterogeneous acceleration system for the large language model.

Another challenge lies in the compilation system, specifically addressing the vast number of operators and optimizing the linkage between them, which is crucial for enhancing computational speed [25]. Current AI chips grapple with the issue of non-uniform data formats for different operators, particularly when dealing with deep learning models. These models usually consist of a series of diverse operators: CNN networks predominantly utilize linear operations such as Matrix Multiplication, Convolution and Pooling operators [2]; while the Transformer networks heavily rely on complex nonlinear operations such as Layer Normalization, Multi-Head Attention Mechanisms (MHA), Embedding Lookups and Positional Encoding [4,5]. Each operator potentially requires distinct data formats to achieve optimal computational efficiency. For instance, convolutional layers might prefer the NHWC format (batch size, height, width, channels) or the NCHW format (batch size, channels, height, width). While in Transformer-like network, the diverse operators will bring more complex data format requirements, especially for those operators in MHA, which requires different operations for different Heads. The absence of a standardized data format in LLM means that when data transitions from one operator to another, it often necessitates format transformations such as reshaping and matrix transposition. These additional operations not only add to the computational load and consume precious computational resources but also introduce extra time lags, thereby reducing the overall system efficiency and throughput.

Besides, in large language models, the operator graph, which defines the computation flow, can be exceedingly complex with hundreds/thousands of operators interconnected in intricate ways. Ensuring that the transition from one operator to another is seamless and efficient becomes a significant challenge. Traditional compilers are designed to optimize code for sequential execution, but the nature of deep learning models requires a different approach. The dataflow graphs of these models present opportunities for parallelism and pipelining that traditional compilers might not fully exploit. Therefore, specialized deep learning compilers and runtime systems have emerged to address these unique requirements. These compilers analyze the operator graph to identify opportunities for optimization, such as operator fusion, which combines adjacent operators into a single, more efficient operation. Another critical aspect is the management of memory access patterns. Efficiently handling the movement of data between different memory hierarchies can significantly impact performance. In summary, the compilation system for large models must be sophisticated enough to handle the complexity of the operator graph, optimize for parallelism, and manage memory access efficiently.

In this work, we develop an efficient CPU-FPGA heterogeneous acceleration system for large language model, where FPGA executes the core computational operators, and CPU executes the dynamic compilation process. When it comes to operational efficiency, our innovative design proposal outperforms both conventional GPU chips and state-of-the-art FPGA systems. Such superior performance is achieved through a combination of optimized hardware architecture, efficient data handling mechanisms, and advanced dynamic compilation scheme.

The main contributions are:

1. We analyzed the computational requirements of FFN (Feed-Forward Network) and MHA (Multi-Head Attention), and proposed high efficiency mix-precision computation unit together with group systolic architecture. Besides, log-scale structured-sparsity together with the block-level quantization method are proposed to balance the hardware efficiency and algorithm accuracy.
2. We analyzed the compilation requirements in LLM and devised unified and universal data format for all of the operator and all of the high-dimensional tensor structures within AI algorithms, enabling the system to execute the operators swiftly and without any data rearrangement.
3. We developed end-to-end compilation scheme, in which dynamic compilation is used for different input token length, and instruction pipeline strategy is used to reduce latency. The scheme can dynamically compile all of the operators and map the whole model on CPU-FPGA heterogeneous system.
4. Finally, the whole design has been successfully deployed on AMD Xilinx VCU128 FPGA. Our result achieves 1.91× higher throughput and 7.55× higher energy efficiency than GPU, and 10~24% higher performance than state-of-the-art FPGA accelerator of FlightLLM.

## II. BACKGROUND AND RELATED WORK

In this section, we will provide a concise overview about the algorithm of Transformers and Large Language Models, along with the advancements in high-performance LLM chips and the high efficiency FPGA accelerator.

### A. Transformers and LLMs

Transformer model was first introduced in 2017 by the famous paper of "Attention is All You Need"[4], which broke through the limitations of traditional Recurrent Neural Networks (RNNs) and Convolutional Neural Networks (CNNs) when dealing with sequential data by adopting the self-attention mechanism. This mechanism allows the model to consider information from the entire sequence while processing any position within it, thereby theoretically achieving superior parallel processing capabilities and capturing long-range dependencies.

In 2018, the GPT (Generative Pre-trained Transformer) series of models from OpenAI, has showcased the enormous potential of language models in generation and understanding. GPT-3, in particular, is noteworthy with its 175 billion parameters, demonstrating an ability to understand and generate high-quality human language [11]. In 2019, a research team from Google introduced the Text-to-Text Transfer Transformer (T5), a unifying framework that reformulates virtually all natural language processing (NLP) tasks as text-to-text transformation problems [26]. Through pre-training and fine-tuning, T5 demonstrated outstanding performance across a spectrum of NLP tasks, further validating the flexibility and robust adaptability of the Transformer architecture [27].

As research progressed, the Transformer architecture began to be applied to multi-modal tasks, such as image captioning and video understanding, achieving effective integration of cross-modal information. The famous models include GPT Series [11], ChatGLM [28], PaLM [29], Sora [30] and so on. These large models demonstrate remarkable abilities in their respective fields, ranging from simple text generation to complex multi-modal understanding and generation, continuously pushing the boundaries of artificial intelligence technology.

*B. AI Chips for Transformers and LLMs*

AI chips that support the operation of large language models typically require high computational power, high memory bandwidth, and an efficient parallel processing architecture. So far, most of current AI accelerator systems, particularly those for large language models, are exclusively based on CPU-GPU architecture, such as NVIDIA's A100 GPU, Google's TPU [31, 32], Intel's Habana Gaudi AI training processor, Graphcore's Intelligent Processing Unit (IPU), and AMD's Instinct MI200 series GPU. These AI chips usually provide abundant computational resources (> 100 TOP/s), high memory bandwidth (>1 TB/s), and high power (>200W) to support the training and inference of large-scale models. To catalyze the pervasive integration of AI across societal and industrial landscapes, a critical imperative emerges: the necessity for AI's seamless deployment on a diverse spectrum of edge-side hardware, transcending the confines of cloud-based accessibility. FPGA implementation offers flexibility in design, allowing for customization and optimization of the model for specific applications, thus is particularly suitable for the edge application. Currently, there have been several advances in such fields.

The work of [33] maps a 172 GFLOP Neural Machine Translation model with mixed-precision representation to a single FPGA board, which is the first work on implementation a real-life end-to-end NMT model to FPGA. But this work is operated on low frequency of 100MHz and shows relative low throughput. [34] proposes the ViA, a novel vision Transformer accelerator architecture based on FPGA, to execute the transformer application efficiently. They obtained nearly 309.6 GOP/s computing performance in the peek on FP16 data type. [20] propose an integer-only accelerator for Transformer, and the average throughput reaches as high as 762.7 GOPs, demonstrating significant acceleration performance improvement when compared with previous state-of-the-art accelerators. But the integer model is not suitable for the large language model.

FlightLLM [16] is the first FPGA based high performance LLM accelerator. By introducing configurable sparse digital signal processor (DSP) chain for various sparsity patterns, and always-on-chip decode scheme with mixed-precision support to enhance memory bandwidth utilization, the FlightLLM achieves 6.0× higher energy efficiency and 1.8× better cost efficiency than the NVIDIA V100S GPU for Llama2-7B models, with 1.2× higher throughput than the NVIDIA A100 GPU during decoding, showing great potential for the FPGA based LLM acceleration system. However, the bandwidth utilization is 65.9%, indicating a relative long pipeline bubbles between data transmission and computation and can be further improved.

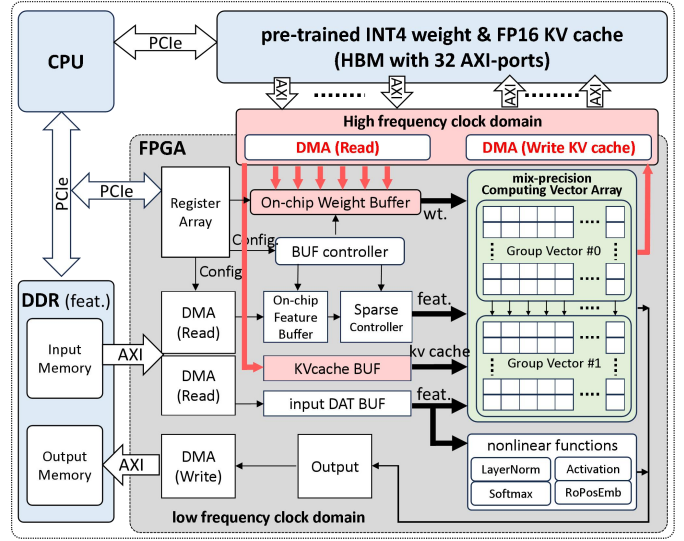

**Fig.2**. Overall architecture of the accelerator proposed in this work, in which "feat." and "kv cache" represent the on-line generated activation data (FP16), "wt." represents the pre-trained weight parameters (INT4).

## III. HARDWARE DESIGN

In this section, we overview the basic architecture of our accelerator and discuss the mix-precision computation unit together with the sparse strategies.

*A. Overall Architecture*

**Fig. 2** provides the basic architectural diagram of the proposed CPU-FPGA Heterogeneous accelerator. The whole system contains four elements: the Central Processing Unit (CPU), High Bandwidth Memory (HBM), DDR memory and the LLM Accelerator. The CPU acts as the brain of the system, interfacing with the remaining components via the Peripheral Component Interconnect Express (PCIe) bus. It wields direct access privileges over DDR and HBM through the AXI-full protocol, enabling both read and write operations. Furthermore, it leverages the AXI lite protocol to interact with the accelerator IP's internal register array, thereby exercising control over the accelerator's operational dynamics.

Within the accelerator IP, various LLM operators have been meticulously incorporated, including MatMUL operator, MHA operator, LayerNorm operator and so on. Among these, MatMUL and MHA operators are connected with HBM (red regions in **Fig. 2**) due to the extremely high speed memory access requirements for weight parameters and KVcaches. When accelerating the MatMUL operators, all of the weight parameters will be pre-processed and stored in HBM, whereas the dynamically generated activation data will be written back to DDR. For the KVcache, a special write Direct Memory Access (DMA) path is constructed to transfer the online generated KVcache into HBM (red lines in **Fig. 2**). Therefore the MHA operator can also take full use of the high bandwidth memory. The other operators, such as the LayerNorm, RMSNorm, Rotary embedding, and nonlinear-activation operators, are connected to the DDR memory with different custom-designed direct memory access (DMA) strategies.

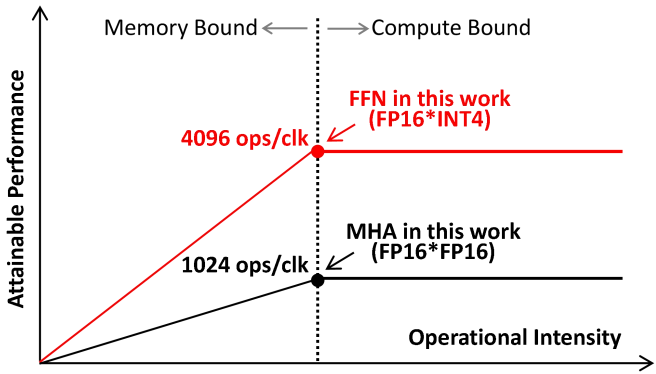

Fig. 3. Roofline model and the designed operation point of this work. Both the FFN operator and MHA operator works at the most effective point in our design. Note, the "ops" represents only multiplication here.

All of the operators are specifically designed with customized DMA modules, tasked with fetching activation data between out-chip DDR and on-chip Block Random Access Memory (BRAM). Meanwhile only the MatMUL and MHA operators are connected to HBM for the high bandwidth weight access. **Fig. 3** depicts the roofline model [35] when accelerating the matrix multiplication (MatMUL) operator, serving as a pivotal analytical tool to contrast the boundaries of memory-limited and computation-limited domains. The operational intensity, quantified as the ratio of teraflops per byte (TFLOP/Bytes), spans the horizontal axis, whereas the vertical axis delineates the achievable performance, measured in teraflops per second (TFLOP/s). The roofline model adeptly demarcates the shift from memory-bound to compute-bound scenarios. This visual juxtaposition offers profound insights into how varying clocking strategies influence the operational intensity and attainable performance of MatMUL, thereby elucidating the intricate interplay between memory bandwidth and computational throughput.

To harness the high-speed communication capability of HBM, we have engineered a computation array that aligns impeccably with the roofline model. In the specific task of LLM operating with a KVcache framework, the MatMUL task can be simplified into vector-matrix multiplication (VMM). Here, the HBM's bandwidth is 8192 bits per cycle (32 AXI ports, and each port transmits 256 bits per cycle). To fully utilize such high bandwidth, we set the system to consume a corresponding 8192 bits of weight parameters per cycle, which means a system computational parallelism of 2048 (under INT4 quantization scheme in FFN). To further increase the computation intensity, we have doubled the computation parallelism, thus the HBM-AXI communication module and DMA module should be operated at a twice higher frequency than that of the computational module. Specifically, we set the computation parallelism to be 4096 in FP16*INT4 based FFN layers (FFN weight consumes 4096*4=16384 bits/cycle), and the parallelism is 1024 in the FP16*FP16 based MHA blocks (MHA KVcache consumes 1024*16==16384 bits/cycle), meanwhile the HBM provides 16384 bits/cycle. In such cases, the system can neither be limited by the memory bound nor by the computation bound.

### B. Design of Mix-Precision Computing PEs

Though LLM model contains a large number of operators, >95% of the computation operations are still MatMUL or VMM. To support the above computation, various systolic arrays have been proposed. Google's TPU[31,32] is the most famous systolic array, which executes the matrix multiplication with a fine-grained pipeline. However, it requires huge numbers of registers to store the temporary data, leading to large dynamic power and circuit area. To increase the efficiency, we have developed the grouped vector systolic computing array (G-VSA)[19], which is shown in **Fig. 4(a)**. Different from TPU-style or broadcast-style systolic array, the input features and weights in G-VSA are transferred into PE array in a row-by-row style, thus the power consumption and area can both be largely decreased.

On the other hand, though the existing LLMs usually quantize the weight parameters to INT4 format, the activation functions in the model remain in FP16 format to ensure the computational accuracy. Therefore, the system requires matrix multiplications in both FP16*INT4 format for the FFN-layers and FP16*FP16 format for the MHA-blocks. Structure of the mix-precision computing PE has been shown in **Fig. 4**, in which $T_{in}$ =128 is the vector length. For the FFN layer, the module can support $T_{in}$=128 number of FP16 feature data * INT4 weight data. While for the MHA block, we set the parallelism to be $T_{in}/4$, therefore the module needs $T_{in}/4$=32 number of FP16 activation data, and $T_{in}/4$=32 number of FP16 KVcache data per clock cycle. Since both the INT4 weight data and FP16 KVcache come from the HBM, the computing PE requires exactly the same bandwidth in the above two cases. Furthermore, we utilized time-division-multiplexing method and share DSPs during each FP16*INT4 computation. By such design, the system can be neither limited by the memory bound nor the computation bound. Finally, to align with the computational requirements of the block-level INT4 quantization algorithm, the unit also supports the multiplicator of Scale value (in FP16 data format) to adjust the final output.

The detailed dataflow in the computation unit has been described in **Fig. 4(b)**. The whole computation can be divided into four steps. Stage-0 is to process all of the inputs. Generally, the FP16 will be split into sign bit (S), exponent bit (E), and mantissa bit (M), while INT4 data will be split into sign bit (S) and mantissa bit (M). Specially in case of MODE-0, each FP16 KVcache data consists of four adjacent INT4 data.

In Stage-1, an exclusive-or (XOR) operation is performed between the sign bits of each FP16 DAT and the INT4 WT. Concurrently, the exponent bits of FP16 data are analyzed by the exponent comparison module, which identifies the highest values among all FP16 inputs. The module computes the exponent distance between each input and the maximum one. Meanwhile, the mantissa bits of FP16 and INT4 will be sent to the multiplier array for a multiplication operation. Within the multipliers, a deliberate design choice has been to preserve the entirety of the mantissa components. This meticulous approach serves to significantly retain the precision of our computational outcomes, ensuring that no fractional detail is lost in the arithmetic processes.

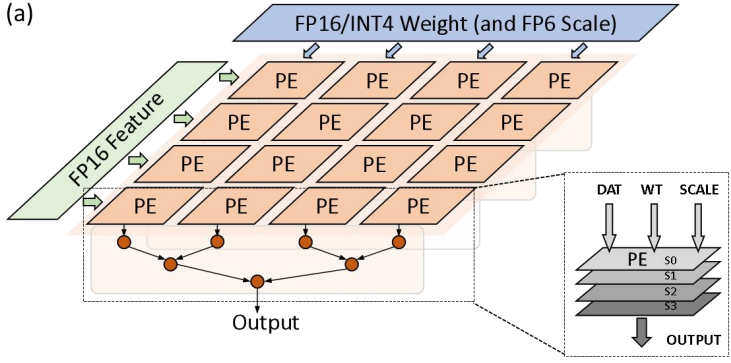

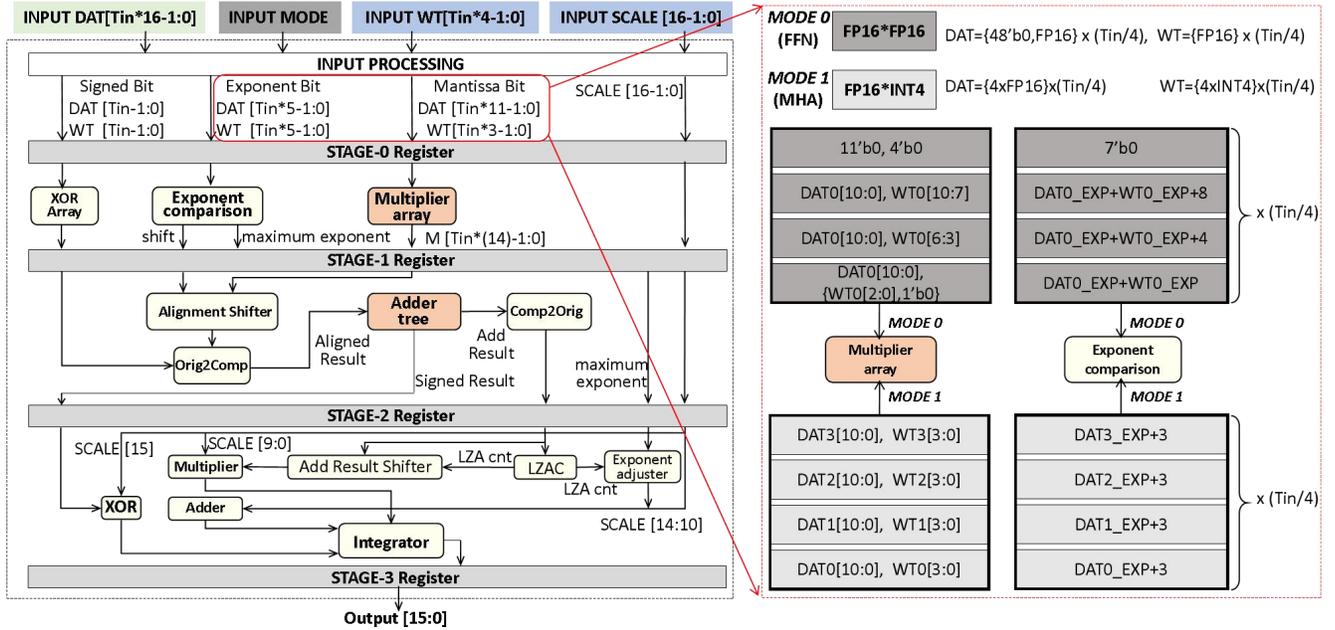

**Fig. 4**. (a) Group vector systolic array. (b) Architecture of the fully customized mix-precision vector multiplier for MatMUL operator, which can support both FP16*INT4 computation for FFN layers and FP16*FP16 computation for MHA layers.

In Stage-2, the exponent distances and the multiplication results obtained in Stage-1 will be fed into the alignment shifter module for aligning the decimal points. Then the shifted mantissa results will be sent into the adder tree module for accumulation. Recognizing that the preliminary shifting operations, the input data width will be increased. To avoid excessive resource consumption, we have specified the bit-width of the adder tree as 19 bits, resulting a balance between resource utilization and computational accuracy.

In Stage-3, the multiplication result and maximum exponent obtained from the previous layers will be converted into specification FP16 data through LZA (Leading Zero Anticipate) module and exponent adjustment module. Then they will be multiplied with the initial input of Scale value and get the final result using floating point multiplication. Finally the result will be input into the integration module to be adjusted to FP16 for output.

To validate the efficiency of our proposed scheme, we design two additional sets of control experiments. In these experiments, a standard pairwise addition-based adder tree was employed; however, the precision of intermediate calculations varied. In baseline-1 case, we configured the intermediate temporary results to be represented in the FP16 format. Conversely, in baseline-2 case, a customized FP20 data format (S1-E6-M13) is used to avoid overflows during computing, meanwhile maintaining high accuracy due to the large bit width of mantissa.

Performance comparison of the different computing unit design methods have been listed in **Table-I**. The proposed mix-precision module exhibits significantly higher efficiency and lower computational errors when compared with the two baseline schemes. Under 100,000 random input tests, our methodology manifests a mere 0.047% error rate on FP16*INT4 mode, and 0.0044% error rate on FP16*FP16 mode. While the baseline method incurs large error rates.

TABLE I. Performance Comparison of the
Mix-Precision Computing Unit using Different Methods

| | | this work | | baseline1 (adder-tree in FP16) | | baseline2 (adder-tree in FP20) | |
|---|---|---|---|---|---|---|---|
| data format | | FP16 * INT4 | FP16 * FP16 | FP16 * INT4 | FP16 * FP16 | FP16 * INT4 | FP16 * FP16 |
| computation error | | 0.0472% | 0.0044% | 2.864% | 14.470% | 2.644% | 0.020% |
| ASIC flow | Total Area | 71664 um² | | 80675 um² +26762 um² | | 110668 um² +30009 um² | |
| | Total Power | 40.34 mW | 10.39 mW | 35.03 mW | 14.66mW | 41.58 mW | 17.90 mW |
| | Maximum Frequency | 1.11 GHz | | 1.03 GHz | 1.28GHz | 1.06 GHz | 1.25GHz |
| FPGA flow | LUT | 24714 | | 24060 + | 6425 | 37320 + | 7870 |
| | FF | 12348 | | 4151 + | 1016 | 4596 + | 1268 |
| | DSP | 128 | | 128 + | 32 | 128 + | 32 |

Besides, our design exhibits the highest comprehensive performance in terms of Power, Performance, and Area (PPA). Firstly, the total area of our design is only 71664 μm², while baseline-1 (with FP16 based adder-tree) is 80,675 + 26,762= 107,437 μm², and that of baseline-2 (with FP20 based adder-tree) is 140,677μm², showing 33.2% area reductions when compared with baseline 1 and 49.1%-decrease with baseline 2. Secondly, our module outperforms competitors in clock frequency, achieving a peak of 1.11GHz on a 28nm CMOS, which is good enough for the practical application. Lastly, concerning power consumption, baseline-1, due to its inherently lower computational complexity in FP16 operations, naturally incurs the least power draw (35.03 + 14.66 = 49.69 mW). Nevertheless, our scheme exhibits only 2% increment in power consumption (40.34mW for FP16*INT4, and 10.39mW for FP16*FP16), demonstrating excellent efficiency. When compared with the baseline-2, our method shows much less consumption power and circuit area. These attributes confirm our approach as the optimal PPA performance.

### C. Further Optimization with Log-scale Sparsity

Large language models often contain billions of parameters. This massive size presents significant challenges in terms of storage, memory, and computational efficiency, especially when deploying these models on devices with limited resources or when attempting to scale up distributed training processes. INT4 quantization, which represents weights using 4-bit integers instead of the typical 32-bit or 16-bit floating-point numbers, offers several advantages: higher memory efficiency, higher computation speed and better bandwidth optimization, thus has been widely used in the inference of LLM. Furthermore, structured sparsity offers significant advantages. It boosts computational speed by skipping zero weights, reduces storage needs for large models, saves energy, and enables complex models to run on devices with limited resources, all while maintaining accuracy.

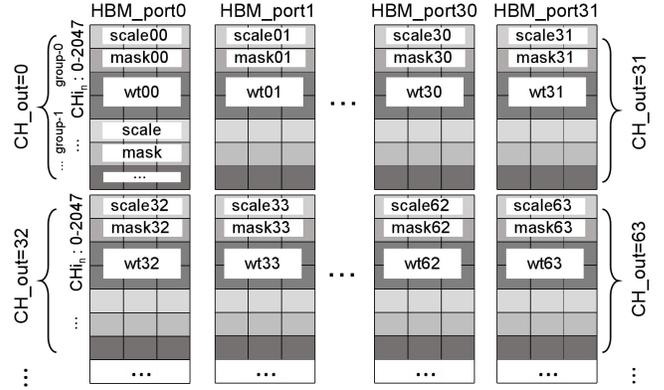

| log-scale structured sparse | case1: dense | case2: 50% | case3: 75% | case4: 87.5% sparsity | |
|---|---|---|---|---|---|
| | | one-hot encoding | addr-in-block | one-hot encoding | addr-in-block |
| CH in one group | 2048 | 2048 | 2048 | 2048 | 2048 |
| scale bits | 256 bits | 256 bits | 256 bits | 256 bits | 256bits |
| mask bits | 0 | 2048 bits | 1536 bits | 2048 bits | 1024 bits |
| wt bits | 8192 bits | 4096 bits | 2048 bits | 1024 bits | 1024 bits |
| Total bits for 2048 CH$_{in}$ (scale + mask + wt) | 8448 bits | 6400 bits | 3840 bits | 3328 bits | 2304 bits |
| effective Bit-width | 4.125 bit | 3.125 bit | 1.875 bit | 1.625 bit | 1.125 bit |
| Performance Enhancement | - | 1.32 × | 2.2 × | 2.54 × | 3.67x |

**Fig.5**. Customized weight parameters package method and analysis of the effective weight bit-width and performance enhancement at different log-scale mix sparsity.

Density bound block (DBB) based structured sparsity has proven to be an effective method, but it is still an open question to explore which kind of structure sparse is the best. Previous work [36] focused on the linear distribution of sparsity (namely 1/8, 2/8, 3/8...), but the hardware utilization is usually low, resulting in low efficiency. The GPU [37] based fix sparsity (2/4 weight sparsity) shows 2× higher efficiency in computation, but its memory access overhead has hardly been reduced. To overcome the above challenges, this paper resents log-scale structured-sparse (namely 1/2, 1/4, 1/8...) accelerator. With the help of time-unrolled micro-architecture, the hardware utilization can reach 100% in a wide range of sparsity, and more importantly, the memory access overhead can be largely decreased.

We utilize structural sparsity into the weight parameters in the matrix multiplication operator. During the execution of matrix multiplication, a broader range of activation data is initially fed into the system. Subsequently, the necessary activation data is selectively picked out based on the mask within the weight parameters. Finally, both the selected activation data and the weight data are jointly directed into the computational array for processing. An efficient sparse DMA has been designed for the feature data.

For the weight data, block quantization [21,22] and hybrid sparsity schemes are used, wherein 128 adjacent parameters are symmetrically quantized and share the same quantization scale parameter. The log-scale mix sparsity is defined as that every group of eight adjacent data blocks must contain at least N zeros, meanwhile the others are non-zeros. For example, when the sparsity is 75%, it means each eight adjacent data blocks must contain at least 6 zeros and at most 2 non-zeros. Based on these analyses, the parameters can be categorized into three types: quantization scales, masks, and the weights themselves, all of which will be stored in HBM. Given that the single HBM AXI-port in this FPGA has a bit width of 256 bits, and each quantization scale is of FP16 data type, the group size will be $256/16 * 128 = 2048$ input channel groups ($CH_{in}$).

**Fig. 5** illustrates the classic weight data shape. Firstly, we pack together all weight parameters belonging to the same $CH_{out}$. For example, all data with $CH_{out}$ indices 0, 32, 64 will be sequentially placed in HBM-AXI-port00, and all data with $CH_{out}$ indices 1, 33, 65 will be sequentially placed in HBM-AXI-port01, and so on. Then, for each weight package, it is divided into $CH_{in}/2048$ portions, and any portion with fewer than 2048 channels will be padded to 2048. Each portion is arranged in the order of scale, mask, and wt.

**Case-1**: when the model is dense, the 2048 $CH_{in}$ groups contain $2048 * 4 = 8192$ bits of INT4 weights, and contains $(2048/128) * 16 = 256$bits of FP16 scale. Therefore the total ratio of scale-to-mask-to-wt will be $256 : 0 : 8192 = 1 : 0 : 32$.

**Case-2**: when layer sparsity of the model is 50%, the 2048 $CH_{in}$ groups contain $2048 * 4 * 50\% = 4096$ bits of INT4 weights. Here we use one-hot encoding scheme for the non-zero mask, such that the 2048 $CH_{in}$ groups contain 2048 bits of mask. And the total ratio of scale-to-mask-to-wt will be $256 : 2048 : 4096 = 1 : 8 : 16$. Note, if the address-in-block scheme (this encoding scheme directly indicates the offset address of each weight) was used, each 2048 $CH_{in}$ group will contain 4096 bits for the mask, which is not efficient here.

**Case-3**: when the layer sparsity is 75%, the one-hot encoding mask requires 2048 bits, while the weight data will be $2048*25\%*4 = 2048$ bits. Consequently, the total ratio of scale-to-mask-to-wt will be $256 : 2048 : 2048 = 1 : 8 : 8$. If address-in-block method was used, the mask would need 1536bits, therefore the total ratio of scale-to-mask-to-wt will be $256 : 1536 : 2048 = 1 : 6 : 8$.

**Case-4**: when the layer sparsity of the model is 87.5%, the one-hot encoding mask requires 2048 bits, and the weight data will be $2048*12.5\%*4 = 1024$ bits. Therefore, the total ratio of scale-to-mask-to-wt will be $256 : 2048 : 1024 = 1 : 8 : 4$. When the sparsity is very high, the cost of the address-in-block method is more efficient. As shown in **Fig. 5**, the 2048 $CH_{in}$ groups contain only 1024 bits of mask. Therefore, the total cost is 2304 bits, and the performance enhancement ratio reaches as high as $3.66\times$ at 87.5% sparsity.

According to the above analysis, the effective weight bit-width is 4.125bit, 3.125bit, 1.875bit, and 1.125bit, respectively. And the performance enhancement ratio will be $1.32\times$, $2.2\times$, $3.67\times$ at different sparsity. In summary, the structured log-scale sparse offers several advantages: higher memory efficiency, higher computation speed and better bandwidth optimization.

**TABLE II. Comparison of Different Sparse Strategies**

| GLM-6B operator | Dense | Sparse strategy-1 | Sparse strategy-2 | Sparse strategy-3 |
|---|---|---|---|---|
| Q | dense, 8.25 MB | dense, 8.25 MB | dense, 8.25MB | dense, 8.25MB |
| K | dense, 0.516 MB | dense, 0.516 MB | dense, 0.516MB | dense, 0.516MB |
| V | dense, 0.516 MB | dense, 0.516 MB | dense, 0.516MB | dense, 0.516MB |
| O | dense, 8.25 MB | 50% sparse, 6.25 MB | 50% sparse, 6.25MB | 50% sparse, 6.25MB |
| h to 4h | dense, 55.23 MB | 50% sparse, 41.8 MB | 75% sparse, 25.08MB | 75% sparse, 25.08MB |
| 4h to h | dense, 27.57 MB | 50%sparse, 20.89 MB | 50%sparse, 20.89 MB | 75%sparse, 12.54MB |
| total wt in a Block | 100.33 MB | 79.22 MB | 61.502 MB | 53.152 MB |
| speedup | 1 × | 1.27 × | 1.63 × | 1.89 × |

| DATABASE | Perplexity | | Zeroshot Accuracy | | | |
|---|---|---|---|---|---|---|
| | WikiText-2 | C4 | Arc-C | Arc-E | BoolQ | Wino. |
| dense | 29.92 | 42.62 | 34.64 | 62.54 | 78.65 | 62.43 |
| Sparse strategy-1 | 38.54 | 49.24 | 32.50 | 60.56 | 76.72 | 56.74 |
| Sparse strategy-2 | 59.24 | 66.4 | 30.54 | 57.78 | 74.98 | 55.88 |
| Sparse strategy-3 | 120.87 | 133.4 | 25.68 | 44.48 | 67.92 | 54.06 |

**Table II** shows the GLM-6B algorithm performance under different sparse strategies. After simplification through mixed-sparsity, the total amount of weight parameters (along with positional encoding parameters) can be significantly reduced, while the algorithm's accuracy remains at a high level [21]. Compared with previous work, our main innovation is that the hardware support mixed-sparsity, allowing us to select different sparsity levels for different operator layers. For example, the QKVO layer can be 50% sparsity, while the h-to-4h layer can be 75% sparsity and so on. Most critically, due to both M and N are powers of two in our design, the hardware computing units can always achieve 100% computational efficiency across different sparsity. Besides, our sparse blocks are larger; for instance, for the same 50% sparsity, GPU uses 2:4, whereas we can use 4:8, 8:16, or 32:64. Although this increases hardware costs, it results in better performance at the algorithmic level. Furthermore, we proposed hybrid encoding scheme to record the position of non-zero weight, in which both one-hot encoding scheme (in low sparsity) and address-in-block scheme (in high sparsity) were used to further improve the efficiency.

## IV. SOFTWARE DESIGN

In this section, we will introduce the compilation strategies aimed at optimally executing the hardware operators.

### A. Unified Data Format based LLM Operator Graph

When executing an artificial intelligence model, the compilation process of the system relies on the operator graph to achieve coherence and efficiency. **Fig. 6** depicts the case of ChatGLM-6B, in which one block will be fused into 17 hardware steps. The key feature is that each operator's inputs and outputs maintain a consistent data structure to achieving optimal execution speed. This consistency allows for smooth transitions from one operator to the next.

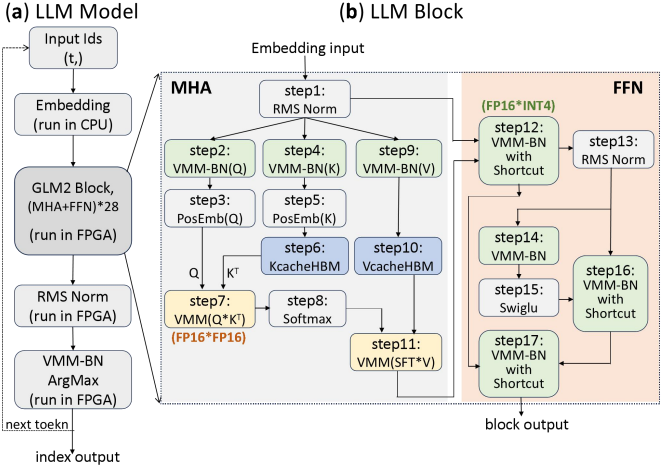

**Fig. 6.** Optimized operator graph, in which the light green boxes represent FP16*INT4 MatMUL, light blue boxes represent HBM-KVcache, and light yellow boxes represent FP16*FP16 MatMUL operators related to KVcache.

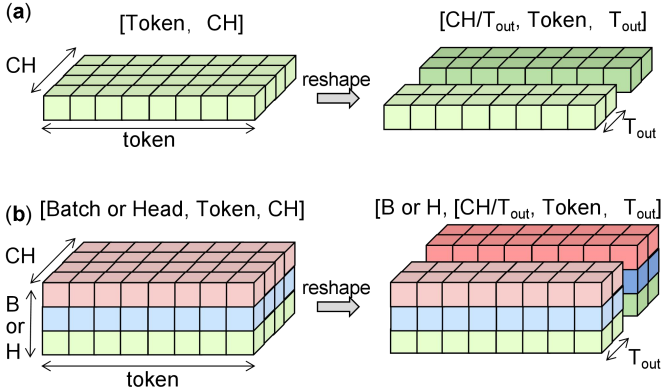

**Fig. 7.** The data packaging method in the accelerator, which ensures all input and output features maintain the same data shape, enables different operators to process the data without any rearrangement.

To maintain a consistent data structure, we developed a universal data parallelism scheme for LLM operators. The scheme is generic and can be adapted to any type of AI algorithm. The text-type input/output activation data innately embodies a two-dimensional structure of (tokens, channels). The data can be easily transformed into three-dimensional tensor representation, namely $[CH/T_{out}, token, T_{out}]$. Here, $T_{out}$ represents the parallelism degree in data channel direction. Similarly, the image-type input/output activation data, which is originally characterized by height (H), width (W), and channels (CH) can be transformed into four-dimensional tensor representation of $[CH/T_{out}, H, W, T_{out}]$. It can be seen that, the text-type and image-type data are sharing with the same tensorization scheme. Such unification not only streamlines data handling but also underscores the inherent compatibility of tensor representations across diverse data modalities.

Moreover, in scenarios where multi-head attention mechanisms introduce an additional head dimension, or batch dimension, the data dimensions can be expanded analogously as $[head\ or\ Batch, CH/T_{out}, H, W, T_{out}]$, ensuring the framework remains versatile. Take the ChatGLM model as an example. The shape of the input is (token, 4096), and weight matrix of $W_q$ is (4096, 4096), of $W_k$ and $W_v$ is (4096, 256). After the matrix multiplication, we will get the three matrixes of Query-Key-Value in (token, 4096), (token, 256), (token, 256), respectively. The Query-Key-Value matrix will be further embedded and reshaped according to the head numbers, generating a three-dimensional matrix of (head, token, 4096/head). Next, the context information will be calculated by $Q*K^T$, in which Q is the Query matrix, $K^T$ is the transpose of Key matrix. Then the attention matrix will be performed by Softmax and finally multiplied by the Value matrix.

It should be noted that K and V are generated by the former MatMUL of K=input*$W_k$ and V=input*$W_v$, respectively. Therefore, both $K^T$ and V can not be pre-treated during the inference process. Besides, the transpose operation breaks the data access pattern for the V matrix, hampering the continuity needed for fine-grained pipelines and necessitating coarse-grained pipelines. Remarkably, the universal data structure introduced in this work is well-suited to solve such problem. Since all of the activation data can be formed in shape of $[CH/T_{out}, token, T_{out}]$, there naturally exists two dimensions of data, namely $[token, T_{out}]$ in a contiguous block of addresses. Using this data structure, we can design a segmented continuous execution of the transpose operation, thus eliminating the need to change the data format.

Another advantage lies in the maximization of efficiency in leveraging AXI burst transfers, which refers to the ability of transferring multiple consecutive data in a single transaction, rather than initiating new transactions. In our design, the data-width in both AXI write-mode and AXI read-mode is fixed at $T_{out}*16$, which is just the same as the smallest data package in $[Batch, CH/T_{out}, H, W, T_{out}]$. Therefore, the incremental address in AXI burst transfer will exactly be the width-dimension or token-dimension in the tensor. This allows for efficient sequential data access, making it particularly useful for reading or writing blocks of data in memory.

### B. Software Architecture and LLM Inference Systems

In addition to the hardware operator design, this work also focuses on software and optimization on the deployment of LLM. **Fig. 8** shows the compilation and deployment of LLM on VCU128 FPGA platform. After sparse and quantized, the LLM will be imported into the compiler, and the compiler will do special optimization including dynamic control and KV-cache, and finally generate instructions, pre-processing weights and runtime control code according to the characteristics of the accelerator to facilitate model deployment. According to the compiled model, we designed a set of LLM inference framework based on LAN (local area network). It uses FPGA and the deployed LLM as the server side, which is responsible for the actual inference of the LLM. Python is used to act as the client to encode and decode the token ids, and conducting direct interaction with users. Furthermore, the appropriate configuration data for the operator registers needs to be meticulously prepared, ensuring that all components are correctly initialized for the upcoming testing phase.

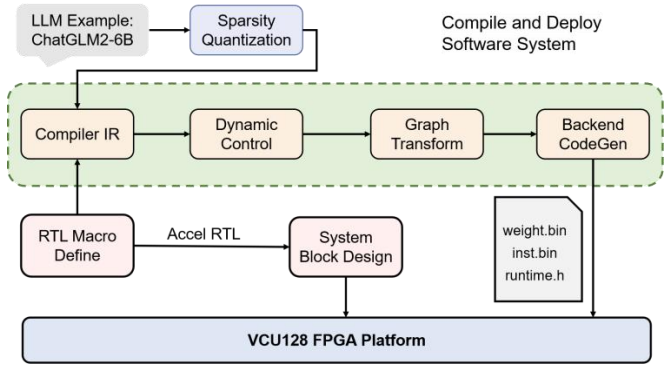

**Fig. 8**. Compilation and deployment of LLM on VCU128 FPGA platform.

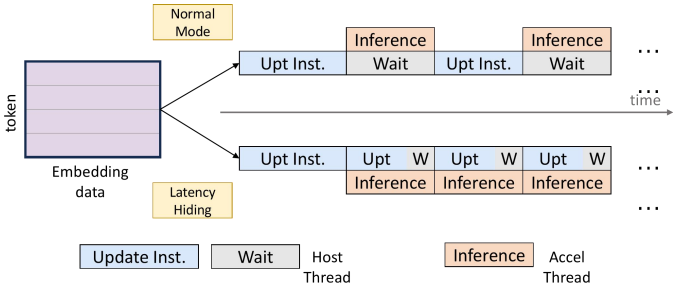

**Fig. 9**. Latency hiding in the inference stage.

In the LLM inference stage, first of all, in order to simplify the dynamic addresses for all operators, the compiler introduces the MAX Token attribute from RTL Macro Define. It replaces the dynamic token in the calculation of the memory space size and address of dynamic data to make the address static, reducing additional computation at runtime. Secondly, hardware instructions also need to adapt to dynamic token. When the compiler evaluates the hardware instructions for the corresponding operator, the token is involved as a Variable and the parameters are recorded as numeric expressions in the form of Directed Acyclic Graph (DAG). If this parameter can be evaluated directly, the compiler returns the result of this instruction, otherwise it is embedded in the runtime code in the form of a simplified code expression for real-time updates. In this approach, the hardware instructions require very little space, making the inference space of KVcache very sufficient.

In addition, since the effective data of all operators after the last attention structure is only on the vector of the last token, the compiler makes additional optimization on this, and gives the actual data offset according to the token parameter. The data of the last token is removed for subsequent calculations to minimize the amount of computation.

Lastly, our design provides pre-configured register mode with auxiliary path to realize the mode operation of instruction pipeline, the compiler can provide the optimization of delay hiding in the inference stage (**Fig. 9**). The accelerator auxiliary mode can encode multiple serialized operator instructions directly from the on-chip DDR to the buffer via the AXI bus. In the host control of the accelerator, only the configuration information of the given serialization instruction needs to be written to the register, such as the address information, the number of valid operators, etc. The accelerator will then complete the rapid start and continuous operation of the serialization instructions. After the accelerator reads the instructions, the host has additional time to perform other calculations before completing the accelerator calculation. As is shown in **Fig. 9**, the delay required for dynamic control instruction updates can be hidden in the time waiting for accelerator calculations to complete. Then, in practical inference, we only need to update the complete instruction before the first model inference, and subsequent instruction updates will be hidden in the last model inference.

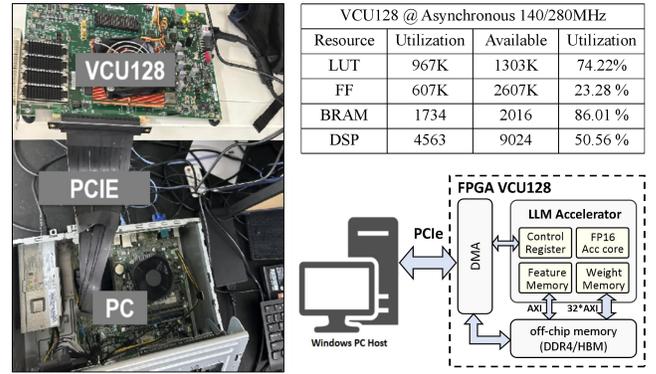

| VCU128 @ Asynchronous 140/280MHz | | | |
|---|---|---|---|
| Resource | Utilization | Available | Utilization |
| LUT | 967K | 1303K | 74.22% |
| FF | 607K | 2607K | 23.28 % |
| BRAM | 1734 | 2016 | 86.01 % |
| DSP | 4563 | 9024 | 50.56 % |

| model | GLM-6B (ref.38) dense | GLM-6B Sparse strategy-1 | GLM-6B Sparse strategy-2 | GLM-6B Sparse strategy-3 |
|---|---|---|---|---|
| average zero-shot accuracy | 59.565 | 56.63 | 54.795 | 48.037 |
| decode speed (token/s) | 52.67 | 66.3 | 77.59 | 85.8 |

**Fig. 10.** Experimental setup for the CPU-FPGA heterogeneous system. The sub paragraph shows the decode speed and zero-shot accuracy on four benchmarks: Arc-C, Arc-E, BoolQ, and Winogrande.(see TABLE-II)

## V. EXPERIMENTAL RESULTS

In this section, we show the experiment results on accelerating LLM modes using AMD Xilinx VCU128 board.

### A. Experiment Setup

The proposed accelerator has been synthesized and implemented using Vivado 2023.1 on AMD Xilinx VCU128 board with 1303K logic elements (LUT), 2607K FF, 9024 DSP blocks, and 2016 Block RAMs on chip. Besides, the FPGA is equipped with 8GB HBM with 460GB/s bandwidth, which is well-suited for the acceleration of GB-level large language models. **Fig. 10** depicts the experimental setup for the system. The host computer (CPU) controls the DDR and HBM via the Peripheral Component Interconnect Express (PCIe) bus. And it leverages the AXI lite protocol to interact with the accelerator IP's internal register array, thereby exercising control over the accelerator's operational dynamics.

The matrix multiplication operator, due to its extremely high demand for handling vast quantities of parameters, is connected with HBM in high frequency of 280MHz. While the other parts and operators work in 140MHz. The operation frequency can be further increased due to timing optimization.

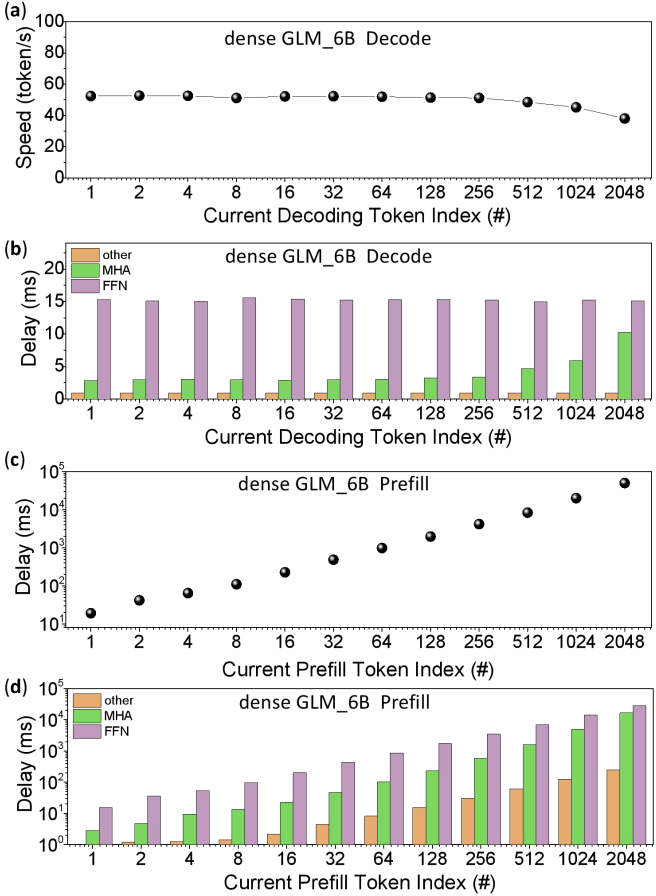

Fig. 11. Acceleration performance on the dense GLM-6B model.

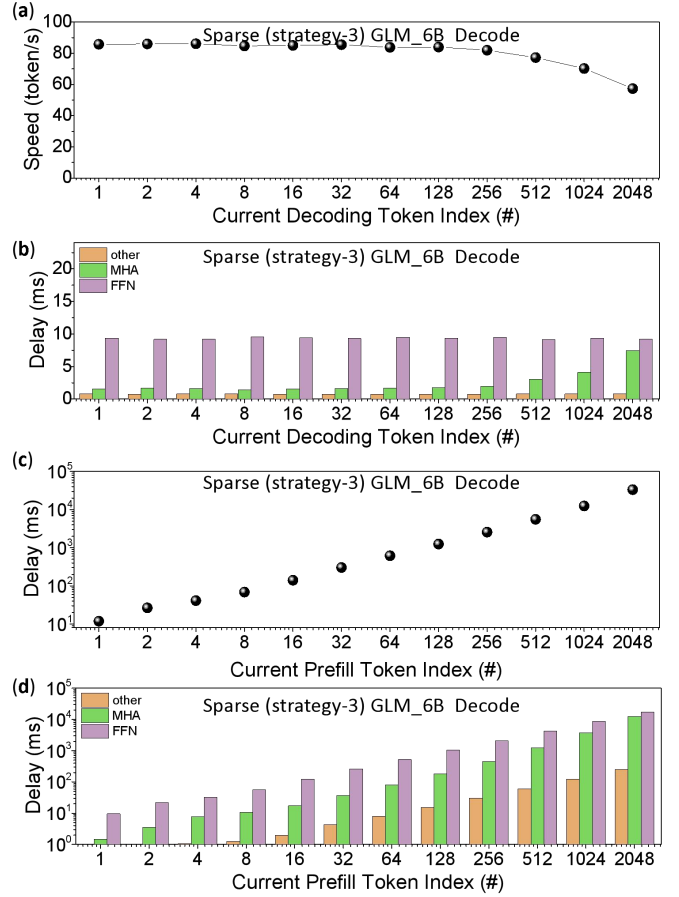

Fig. 12. Acceleration performance on the sparse GLM-6B model.

The sub-paragraph in **Fig. 10** exhibits the experiment results on the acceleration GLM-6B, in which different sparse strategies have been considered. For the original model provided by the official source [38], our chip computation has almost no accuracy loss. After sparsity optimization, the accuracy has been slightly decreased, but the model's speed has increased by approximately 63%. Besides, we have also measured the Qwen-7B LLM model [39]. The Qwen decoding speed (42.5 token/s ~ 69.4 token/s) is slower than that of GLM, which is not only because of the less FFN parameters, but also comes from the highly shared weight-heads in MHA blocks.

### B. Analysis on the Hardware Efficiency

Efficiency of HBM bandwidth utilization is a key point in AI chips. The efficiency can be calculated by the division of real_operation_time and ideal_operation_time. Take the GLM model as an example, the shape of input is (token, 4096), and weight matrix of $W_q$ is (4096, 4096). In the decode stage, the input token will always be 1, so the ideal_operation_time will be 4096*4096*4bit / (8192 bit/cycle) * 3.571ns = 29.25μs. Meanwhile the real_operation_time can be directly measured to be 38.5μs, therefore the HBM bandwidth utilization is 29.25μs/38.5μs = 75.97%. Similarly, we can calculate the utilization of each MatMUL layer and finally observe that they all lie between 70% and 80%, and the average is about 75%.

**Fig. 11** shows the performance on the acceleration of GLM-6B model. The decode speed remains relatively stable (~90 token/s) when the decoding token is less than 512. **Fig. 11(b)** shows the detailed latency breakdown. The whole model can be divided into three parts, namely multi-head attention (MHA), feed forward networks (FFN), and the others. The runtime of FFN is independent to the decode length, but the runtime of MHA has a quadratic relationship with input token length. With the token increasing, the MHA latency will gradually dominate the system. **Fig. 11(c)** and **Fig. 11(d)** shows the prefill runtime, respectively. Due to the increase of token length in prefill, the corresponding computation latency will increase proportionally. In such cases, the bottleneck of the system will be the computation throughput, rather than the data access. **Fig. 12** shows the performance on the acceleration of sparse GLM model, the first decode delay is only 10.8 ms, corresponding a peak speed about 85.8 token/s at a power of 56.86W, showing 1.91× higher throughput and 7.55× higher energy efficiency than GPU [38].

We also measured the Qwen-7B model[39], and all results are quite similar to that of GLM-6B. In Qwen-7B model, the decoding speed using sparse strategy-3 is 69.4token/s, which is slower than GLM-6B due to the larger number of VMM parameters (7B vs. 6B), and more KV heads (4 vs. 2) in MHA block with longer latency in KVcache parameters loading.

## TABLE III. Evaluation of EdgeLLM on DDR system

| delay (μs) test case<br>Dense GLM operator | | Decode token=128 | | Prefill token=128 | |
|---|---|---|---|---|---|
| | | HBM | vs DDR | HBM | vs DDR |
| STEP 1 | LayerNorm | 9.55 | 15.84 | 533.35 | 694.86 |
| STEP 2 | VMM-BN(Q) | 47.12 | 181.66 | 4770.07 | 7840.94 |
| STEP 3 | EMB_Q | 7.79 | 13.70 | 274.29 | 351.03 |
| STEP 4 | VMM-BN(K) | 2.15 | 12.61 | 476.38 | 649.70 |
| STEP 5 | EMB_K | 0.44 | 1.57 | 24.99 | 33.15 |
| STEP 6 | DAT2HBM | 0.23 | 1.63 | 70.42 | 36.46 |
| STEP 7 | TRP | 5.83 | 10.06 | 672.66 | 837.16 |
| STEP 8 | SOFTMAX | 43.38 | 48.68 | 872.54 | 1048.91 |
| STEP 9 | VMM-BN(V) | 1.97 | 10.72 | 475.36 | 650.17 |
| STEP 10 | DAT2HBM | 0.29 | 2.23 | 69.95 | 35.44 |
| STEP 11 | F2W | 5.73 | 9.64 | 614.95 | 837.49 |
| STEP 12 | VMMBNRES0 | 48.34 | 177.30 | 4725.42 | 7845.11 |
| STEP 13 | LayerNorm | 9.52 | 14.48 | 533.76 | 694.53 |
| STEP 14 | VMMBN1 | 137.98 | 596.56 | 16063.43 | 26306.36 |
| STEP 15 | ACT | 15.36 | 33.83 | 890.43 | 1142.23 |
| STEP 16 | VMMBNRES1 | 143.98 | 594.59 | 16007.04 | 26319.11 |
| STEP 17 | VMMBNRES2 | 191.68 | 707.03 | 23429.09 | 75931.96 |
| STEP 18 | Outlayer_LN | 9.75 | 14.40 | 19.86 | 23.09 |
| STEP 19 | VMMBN_Arg | 648.81 | 2759.7 | 639.63 | 2762.25 |

| Summary | Decode token=128 HBM vs DDR | | Prefill token=128 HBM vs DDR | |
|---|---|---|---|---|
| (step1 ~ step17) single block delay (ms) | 671.10 | 2432.12 | 70504.12 | 151254.59 |
| (step1 ~ step17)*28 + step18 + step19 Total LLM delay (ms) | 19449.23 | 70873.4 | 1974774.7 | 4237913.9 |
| speed (token/s) | 51.42 | 14.11 | 0.51 | 0.24 |

## TABLE IV. Energy Consumption of Different Operators

| operator type | power (W) @140/280MHz | operator type | power (W) @140/280MHz |
|---|---|---|---|
| standby | 40.36 | step9: VMM-BN(V) | 42.84 |
| step1: RMS Norm | 41.02 | step10: VcacheHBM | 40.62 |
| step2: VMM-BN(Q) | 54.02 | step11: VMM(SFT*V) | 40.92 |
| step3: PosEmb(Q) | 40.81 | step12: VMM-BN-RES | 57.25 |
| step4: VMM-BN(K) | 42.79 | step13: RMS Norm | 40.97 |
| step5: PosEmb(K) | 40.63 | step14: VMM-BN | 55.13 |
| step6: KcacheHBM | 40.62 | step15: Swiglu | 41.11 |
| step7: VMM(Q*K$^T$) | 41.01 | step16: VMM-BN-Res | 58.13 |
| step8: Softmax | 40.65 | step17: VMM-BN-Res | 53.23 |
| **Normalized Average power : 56.86W** | | | |

## TABLE V. Efficiency Comparison on Different Platforms

| Platform | A100 GPU | FlightLLM [1] | | EdgeLLM @ VCU128 (this work) | |
|---|---|---|---|---|---|
| | | FPGA-U280 | FPGA-VHK158 | | |
| Bandwidth Utilization | ~30% | 65.9% | 64.8% | ~75% | |
| Throughput | ~45 token/s | ~55token/s (on 7B LLM) | 92.5 token/s (on 7B LLM) | 85.8 token/s (on 6B LLM) | 69.4 token/s (on 7B LLM) |
| Power | ~220W | 45W | 155 W | 56.8W | |
| energy efficiency | 0.2 token/J | 1.22 token/J | 0.6 token/J | 1.51 token/J | 1.23 token/J |

To demonstrate the high efficiency of our EdgeLLM framework, we have done additional experiments and evaluated our accelerator's performance on a non-HBM system, where DDR is used instead. In edge system, DDR is the most common seen memory chips with about 60 GB/s bandwidth. As shown in the **TABLE III**, we present both decoding and prefill test cases. It can be observed that, in decode mode, the primary bottleneck of the DDR system lies in several operators related to Matrix-Vector Multiplication. This results in the token generation speed in decoding stage being only about 25% of that of the HBM system. In prefill stage, due to the reusability of weight parameters, which reduces the system's demand for communication bandwidth, the speed reduction is relatively lower. And further experiments show that if the prefill token length was larger, the performance decrease of DDR system will become even smaller. Overall, though the token generation speed in both prefill and decode stage will be decreased in pure DDR system, the permanence of our EdgeLLM is still good enough for edge applications.

**TABLE IV** shows the energy consumption when accelerating each single operator. Overall, the normalized average of the FPGA is 56.86W at 140/280MHz, in which the standby power (after loading the 140/280MHz bitstream) is 40.36 W. It can be seen that most of the nonlinear function (such as LayerNorm/RMSNorm, Rotary Embedding, Softmax) consumes about 41 W, corresponding to a net power to be only ~0.7W. While the matrix-multiplication based operators show relative higher power consumption ranges from 2W to 18W, mainly due to the high bandwidth of HBM workload and intensive computations.

**Table V** shows the efficiency comparison on different platforms. GPU is widely used in LLMs. When accelerating large batch size data (typically for cloud scenarios), their performance and efficiency are very high. However, when the batch size is 1 (typically for edge application scenarios), the efficiency becomes typically lower than 30%. Besides, the energy consumption of GPU is also very high. FlightLLM [16] is currently state-of-the-art FPGA accelerator, enabling efficient LLMs inference with ~66% bandwidth utilization and 6× better energy efficiency than GPU. Attributable to the synergistic effects, our integrated solution showcases enhanced performance over the FlightLLM by 11%-higher bandwidth utilization and up to 24% higher energy efficiency, showing great potential for the practical edge-AI application.

Currently, almost all of the LLMs follow similar operator structures: MHA+FFN. We have implemented comprehensive optimizations on the above operators such that our framework has great general applicability and can be easily to be applied to other types of LLMs besides GLM and Qwen. However, based on our analysis, the potential limitation is the RotaryEmbedding operator, which usually varies at different LLM models. To execute such operator, two methods can be used: one is the custom-designed operator for different models, and the other one is using CPU + ElementWiseOperator. The former one shows high hardware efficiency for specific model but can not be directly used for any other model, while the later one shows higher general applicability but with low speed. Besides, all operators in this proposal are executed in a temporal-mode, with one operator starting only after the previous one has finished. Future optimizations could explore the parallel execution of different operators.

## VI. Conclusion

This study proposes EdgeLLM, a high-performance CPU-FPGA heterogeneous acceleration scheme for LLM. In terms of hardware, we analyzed the computational requirements of FFN and MHA, and proposed high efficiency mix-precision computation unit together with group systolic architecture. Meanwhile specific optimization on log-scale structured weight sparsity was proposed, though the algorithm accuracy has been slightly decreased, the hardware efficiency can be largely increased. In terms of software, we analyzed the whole operator graph, and devised a unified and universal data format for all of the operators within AI algorithms, enabling different operators to process without any data rearrangement. Then we propose end-to-end compilation scheme that can dynamically compile all of the operators and map the whole model. This solution has good versatility and can be adapted to multiple large model algorithms. For instance, the accelerator has been successfully deployed on AMD Xilinx VCU128 FPGA platform. Our result achieves 1.91× higher throughput and 7.55× higher energy efficiency than GPU NVIDIA A100-SXM4-80G, and shows about 10~24% higher performance than state-of-the-art FPGA accelerator of FlightLLM. Overall, EdgeLLM addresses the issues of intensive computational requirements, heavy memory access, diverse operator types, difficulties in compilation of LLMs, showing great potential for the practical application in edge scenarios (such as robots) in the near future.

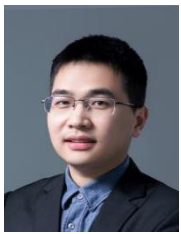
Mingqiang Huang received the B.S. and Ph.D. degrees in physics from Huazhong University of Science and Technology, Wuhan, China, in 2013 and 2018, respectively. From 2018 to 2019, he was a Research Fellow with Nanyang Technological University, Singapore. Since November 2019, he has been with Shenzhen Institute of Advanced Technology, Chinese Academy of Sciences as a Research Associate Professor. His current research interests include energy-efficient computing, artificial intelligence (AI) hardware accelerators.

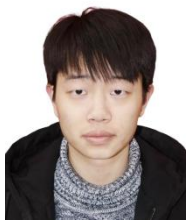
Ao Shen is currently pursuing the master′s degree in microelectronics science and engineering with the School of Microelectronics, Southern University of Science and Technology, Shenzhen, China. His research interests include neural network compiler and digital system design.

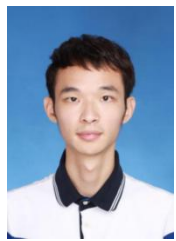
Kai Li is currently the Ph.D candidate in microelectronics science and engineering with the School of Microelectronics, the Southern University of Science and Technology, Shenzhen, China. His research interests include deep learning, and neural network acceleration.

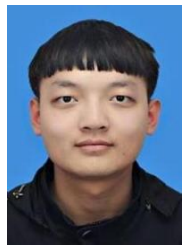
HaoXiang Peng is currently the postgraduate student in microelectronics science and engineering with the School of Microelectronics, the Southern University of Science and Technology, Shenzhen, China. His research interests include deep learning, digital system design, and neural network acceleration.

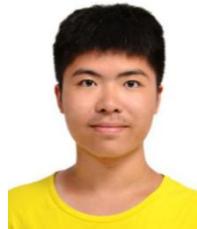
Boyu Li received the B.S. degree from the School of Microelectornics, Southern University of Science and Technology, Shenzhen, China, in 2022. He is currently pursuing the Ph.D. degree with the Department of Electrical and Electronic Engineering, The University of Hong Kong, Hong Kong. His current research interests include deep learning, digital system design, and neural network acceleration.

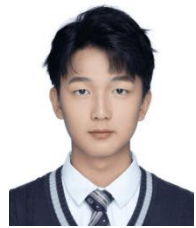
Yupeng Su is a senior student from School of Microelectronics in Southern University of Science and Technology, Shenzhen, China. His research interests includes Efficient and Low-resource Methods for NLP, Model Deployment on Edge, and AI Accelerators Design.

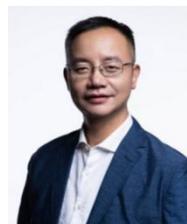
Hao Yu received the Ph.D. degree in electrical engineering from the University of California, Los Angeles, USA, in 2007. He is currently with the Southern University of Science and Technology, Shenzhen, China. His research interests include energy-efficient data links, sensors, and analysis. Dr. Yu has about 282 perreviewed IEEE/ACM publications; three best paper award nominations for Design Automation Conference in 2006, International Conference on Computer-Aided Design in 2006, and international conference on VLSI design automation in Asia and South Pacific region in 2012; He is an Associate Editor for NatureSciReports, IEEE Transaction on Biomedical Circuits and Systems, ACM Transactions on Embedded Computing Systems, Elsevier Microelectronics, etc., and a Technical Program Committee Member for IEEE Custom Integrated Circuits Conference, IEEE Asian Solid-State Circuits Conference, ACM-DAC, ACM Design, Automation and Test in Europe Conference and Exhibition, etc., and of many IEEE/ACM international journals and conferences. He is a senior member of IEEE and ACM.